\documentclass[%
 reprint,
 amsmath,amssymb,
 aps,
]{revtex4-2}
\usepackage{float}
\usepackage{graphicx}% Include figure files
\usepackage{dcolumn}% Align table columns on decimal point
\usepackage{bm}% bold math
\begin{document}
\textit{\textbf{}}% ****** Start of file apssamp.tex ******
\preprint{APS/123-QED}

\title{Generalized Misner-Sharp energy in $f(R,\mathcal{G})$ gravity}% Force line breaks with \\

\author{Amin Rezaei Akbarieh}\email{am.rezaei@tabrizu.ac.ir}
\author{Navid Safarzadeh Ilkhchi}\email{navidsafarzadeh1402@ms.tabrizu.ac.ir}
\affiliation{%
Faculty of Physics, University of Tabriz, Tabriz, Iran
}%


\author{Yaghoub Heydarzade}\thanks{yheydarzade@bilkent.edu.tr}
\affiliation{
 Department of Mathematics, Faculty of Sciences, Bilkent University, 06800 Ankara, Turkey
}%
%\affiliation{ Third institution, the second for Charlie Author }%
%\affiliation{% Authors' institution and/or address\\ This line break forced with \textbackslash\textbackslash }%


\date{\today}% It is always \today, today,
\begin{abstract}
In this work, we explore the formulation of the Misner-Sharp energy within the framework of $f(R, \mathcal{G})$ gravity, a modified theory incorporating the Ricci scalar $R$ and the Gauss-Bonnet scalar $\mathcal{G}$. By extending the quasilocal energy definition to both static spherically symmetric spacetime and the dynamic Friedmann-Lemaitre-Robertson-Walker (FLRW) spacetime, we derive explicit expressions for the generalized Misner-Sharp energy using two complementary approaches: the integration method and the conserved charge method based on the Kodama vector. Our analysis shows that the Misner-Sharp energy expression in $f(R, \mathcal{G})$ gravity reduces to standard $f(R)$ gravity results when the Gauss-Bonnet term is absent, revealing how curvature modifications influence the geometric structure and dynamics of cosmic evolution. Furthermore, we investigate the thermodynamic properties at the apparent horizon associated with the FLRW background, and we find a connection to non-equilibrium thermodynamics unique to $f(R, \mathcal{G})$ gravity. These findings underscore the subtle and fundamental role of curvature corrections in determining the energy distribution and thermodynamic behavior of gravitational systems.
\end{abstract}

%\keywords{Suggested keywords}%Use show keys class option if keyword
\maketitle
\section{Introduction}
Energy is a fundamental concept in physics, yet defining it within the context of gravitation has long presented challenges. In everyday physics, energy is relatively simple—consider a ball rolling down a hill or the warmth generated by a fire. However, when it comes to gravity, the situation becomes more complicated. General relativity, Einstein’s groundbreaking theory, reveals that gravity is not merely a force, but a curvature of spacetime induced by mass and energy \cite{Mie:1913ice,Wald:1984rg}. This curvature complicates our understanding of energy, particularly since the strong equivalence principle suggests that gravity can locally vanish in specific coordinates \cite{Karch:2000ct,Bazeia:2004yw,Oda:2001ss,tHooft:2008mxs}, resulting in no clear way to ascertain a local energy density for the gravitational field itself \cite{DeFelice:2010aj,Misner:1964je}. Consequently, physicists have shifted their focus to broader concepts, examining energy across entire regions of spacetime rather than at local points \cite{Brown:1992br}.\\
One of the most elegant solutions to this problem is the Misner-Sharp energy, introduced by Charles Misner and David Sharp   in 1964 \cite{Misner:1964je}. This quasilocal energy is specifically designed for systems exhibiting spherical symmetry, such as black holes or expanding universes. It serves as a powerful tool because it captures the total energy contained within a sphere, encompassing contributions from matter, radiation, and the gravitational field itself. In the context of Einstein's gravity, the Misner-Sharp energy takes a straightforward form: 
\begin{eqnarray}\label{eq0}
     E(r) = \frac{r}{2G} \left(1 - h^{ab} \partial_a r \partial_b r \right), 
\end{eqnarray}
where $r$ denotes the radius of the sphere, $G$ represents the Newton's gravitational constant and $h^{ab}$ refers to the inverse metric of the two-dimensional surface. This formulation elegantly connects to physical phenomena: It yields the Schwarzschild mass for black holes \cite{Nielsen:2008kd}, it reduces the Newtonian mass in weak gravitational fields \cite{Moffat:2004bm}, and even relates to the thermodynamics of horizons through the first law \cite{Hayward:1997jp,Hayward:1998ee,Cai:2005ra}. Its versatility makes it a preferred approach for exploring gravitational dynamics.\\
Over the years, researchers have further developed this concept by investigating how the Misner-Sharp energy behaves beyond the framework of Einstein's theory \cite{Hayward:1993wb}. In modified gravity theories, which adjust the principles of general relativity to address cosmic phenomena such as the accelerating expansion of the universe, the Misner-Sharp energy has been adapted to new frameworks. Maeda and Nozawa (2008) investigated this energy in Gauss-Bonnet gravity, incorporating higher-order curvature terms into the formulation \cite{Maeda:2007uu}. Subsequently, researchers (2009) analyzed the Misner-Sharp energy in the context of f(R) gravity, where the Ricci scalar R in the gravitational action is replaced by a function $f(R)$ \cite{Cai:2009qf}. These studies demonstrate that the Misner-Sharp energy remains relevant in modified theories. However, it acquires additional terms that reflect the new physics, providing insight into how these theories transform our understanding of energy and spacetime.\\
Among the many extensions of General Relativity, $f(R,\mathcal{G})$ gravity is an especially fascinating approach. What makes $f(R, \mathcal{G})$ gravity important is its particular ability to explain the accelerating expansion of the universe without relying on dark energy \cite{Zhou:2009cy}. These kinds of theories have been extensively studied in connection with frameworks such as string theory and higher-dimensional gravity models \cite{Greene:2015fva,Emparan:2008eg}. Many studies have explored how these models can account for both the rapid inflation of the early universe and its later acceleration, showing that they offer realistic possibilities for understanding cosmic history. Researchers have investigated extensively the field equations in this theory and how they shape the development of the universe. Recent work \cite{Gurses:2024tka}, suggests that the extra curvature effects in modified gravity theories can act like a geometric perfect fluid \cite{Capozziello:2019wfi,Gurses:2020kpv}. This perspective implies that the universe’s late-time acceleration might come entirely from spacetime’s curvature, with no need for dark energy \cite{Planck:2018vyg}. Some studies explore quadratic gravity equations and illustrate how this geometric fluid naturally arises within these frameworks \cite{Gurses:2024tka,Deser:2002rt,Planck:2018vyg}. Even with these advances, some parts of $f(R, \mathcal{G})$ gravity still need more attention. One example is the Misner-Sharp energy concept within this framework. This idea is vital for understanding gravitational thermodynamics and local energy in General Relativity and other alternative theories, but its role in $f(R,\mathcal{G})$ gravity isn’t fully clear yet. Exploring this gap could lead to a better understanding of spacetime’s energy and causal structure in this modified theory.\\
In this paper, we address the challenge of generalizing the Misner-Sharp energy within the framework of $f(R, \mathcal{G})$ gravity, exploring its implications for energy and thermodynamics. Our focus is centered on three primary scenarios: a general spherically symmetric spacetime, the static case (such as black holes), and the FLRW spacetime. Employing two distinct approaches—the integration method and the conserved charge method utilizing the Kodama vector—we derive coherent expressions for this energy. We investigate its behavior in the FLRW universe, relating it to the total matter energy contained within a sphere, while also delving into the thermodynamics at the apparent horizon. Our findings illuminate how curvature corrections in $f(R, \mathcal{G})$ gravity influence energy distribution and suggest deeper connections to the evolution of the universe.\\
The structure of this paper is as follows: In Section II, we compute the Misner-Sharp energy for $f(R, \mathcal{G})$ gravity using two distinct approaches: the integral method and the conserved charge method. In Section III, we evaluate the Misner-Sharp energy in spherical configurations, divided into two subsections: the static spherically symmetric case and the  dynamical FLRW case. In Section IV, we investigate the thermodynamics in an FLRW background, utilizing the Misner-Sharp energy. Finally, in Section V we present the conclusions of the study.\\

%\tableofcontents
\section{GENERALIZED MISNER-SHARP ENERGY IN $f(R,\mathcal{G})$ THEORY: GENERAL CASE}
Here, we shall derive Misner-Sharp energy for the \textit{$f(R,\mathcal{G})$} gravity model using two different approaches and show that both methods yield the same result.
\subsection{Integration method}
We begin by deriving the Misner-Sharp energy using the integration method. The gravitational action of \textit{$f(R,\mathcal{G})$} is given by
\begin{eqnarray}\label{eq000}
S = \frac{1}{16 \pi G} \int d^4x \sqrt{-g} f(R,\mathcal{G}) + S_{\text{m}},
\end{eqnarray}
where, \(\textit{G}\) represents Newton's gravitational constant, \(\textit{g}\) is the determinant of the metric, \(S_{\text{m}}\) denotes the action for matter fields, and \(\mathcal{G}\) is defined as
\begin{eqnarray}
    \mathcal{G} \equiv R^2 - 4R_{\alpha\beta}R^{\alpha\beta} + R_{\alpha\beta\gamma\delta}R^{\alpha\beta\gamma\delta},
\end{eqnarray}
where, $R$ is the Ricci scalar, $R_{\alpha\beta}$ is the Ricci tensor, and $R_{\alpha\beta\gamma\delta}$ is the Riemann curvature tensor.
\begin{widetext}
By varying the gravitational action with respect to the metric tensor $g_{\mu\nu}$, we obtains the field equations
\begin{eqnarray}\label{eq002}
G_{\mu\nu}&\equiv&R_{\mu\nu}-\frac{1}{2}g_{\mu\nu}R\nonumber\\
&=& \frac{1}{f_R}\Big[\nabla_\mu \nabla_\nu f_R - g_{\mu\nu}\Box f_R +2R\nabla_\mu \nabla_\nu f_{\mathcal{G}}- 2g_{\mu\nu} R\;\Box f_{\mathcal{G}} - 4R_\mu^\lambda \nabla_\lambda\nabla_\nu f_{\mathcal{G}} -4R_\nu^\lambda\nabla_\lambda\nabla_\mu f_{\mathcal{G}} +4 R_{\mu\nu}\Box f_{\mathcal{G}}\nonumber \\
&& + 4g_{\mu\nu}R^{\alpha\beta}\nabla_\alpha \nabla_\beta f_{\mathcal{G}} +4R_{\mu\alpha\beta\nu} \nabla^\alpha\nabla^\beta f_{\mathcal{G}} -\frac{1}{2}g_{\mu\nu}\left(Rf_R+\mathcal{G}f_{\mathcal{G}}-f\right)\Big]+\frac{8\pi G}{f_R}T_{\mu\nu},
\end{eqnarray}
where $f_R=df(R,\mathcal{G})/dR$, $f_{\mathcal{G}}=df(R,\mathcal{G})/d\mathcal{G}$, and $T_{\mu\nu}$ is the energy-momentum tensor. 
\\We consider a four dimensional spherical symmetric spacetime with the line element in double null coordinates as
\begin{eqnarray}\label{eq001}
ds^2 = -2e^{-\varphi(u,v)} du dv + r^2(u, v) (d\theta^2 + \sin^2\theta \, d\phi^2).
\end{eqnarray} 
The non-zero components of the field equations  (\ref{eq002}) for the metric (\ref{eq001}) are given by
\begin{eqnarray}\label{eq008}
8 \pi G T_{uu} &=& -\frac{1}{r^2} \Big[ 4f_{{\mathcal{G}},uu} + r^2 \left(f_{R,u} \varphi_{,u} + f_{R,uu}  \right) + 2f_{R} r\left( r_{,u} \varphi_{,u} + r_{,uu}  \right) + 8e^{\varphi} r_{,u} \left( r_{,v} f_{{\mathcal{G}},uu}+ f_{{\mathcal{G}},v} \left( r_{,u} \varphi_{,u} + r_{,uu}  \right)  \right)\nonumber\\
&&+ 4f_{{\mathcal{G}},u} \left(\varphi_{,u}+ 2e^{\varphi} r_{,v} \left( 2r_{,u} \varphi_{,u} + r_{,uu} \right)  \right)  \Big],
\notag \\
8 \pi G T_{vv} &=& -\frac{1}{r^2} \Big[ 4f_{{\mathcal{G}},vv} + r^2 \left( f_{R,v} \varphi_{,v} + f_{R,vv}  \right)+ 2f_{R} r\left( r_{,v} \varphi_{,v} + r_{,vv}  \right) + 8e^{\varphi} r_{,v} \left( r_{,u} f_{{\mathcal{G}},vv} + f_{{\mathcal{G}},u} \left( r_{,v} \varphi_{,v} + r_{,vv}  \right)  \right) \nonumber\\
&&+ 4f_{{\mathcal{G}},v} \left(\varphi_{,v}+ 2e^{\varphi} r_{,u} ( 2r_{,v} \varphi_{,v} + r_{,vv} )  \right)  \Big],
\notag \\
8 \pi G T_{uv} &=& \frac{1}{2} e^{-\varphi} f + f_{R,uv} + \frac{2}{r} \left( r_{,v} f_{R,u} + f_{R,v} r_{,u} - f_{R} r_{,uv} \right) + f_{R} \varphi_{,uv} + \frac{4}{r^2} \Big\{ f_{{\mathcal{G}},uv} + f_{{\mathcal{G}}} \varphi_{,uv} + 2e^{\varphi} \Big[ f_{{\mathcal{G}},v} r_{,u} r_{,uv}\nonumber\\
&&+ r_{,v} \left( r_{,u} f_{{\mathcal{G}},uv} + f_{{\mathcal{G}},u} r_{,uv} \right) + f_{{\mathcal{G}}} \Big( r_{,vv} \left( r_{,u} \varphi_{,u} + r_{,uu} \right)-r_{,uv}^2+ r_{,v} \left( r_{,u} \left( \varphi_{,v} \varphi_{,u} + \varphi_{,uv} \right) + \varphi_{,v} r_{,uu} \right) \Big)\Big] \Big\}.
\end{eqnarray}
\end{widetext}
In Einstein's theory of gravity, the Misner-Sharp energy within a radius $r$ is given by (\ref{eq0}). This energy not only describes physical phenomena in general relativity, such as the Schwarzschild energy for black holes and the Newtonian mass in the weak field limit, but also plays a crucial role in connecting Einstein’s equations to thermodynamic laws. The generalized form of the Misner-Sharp energy allows us to express the gravitational field equations in the form of the first law of thermodynamics
\begin{eqnarray}
dE = A \Psi_a dx^a + W dV,
\end{eqnarray}
where $A=4\pi r^2$ represents the area of the sphere radius r, $W=-\frac{1}{2}h^{ab}T_{ab}$ is the work density, \(\Psi_a\) is the energy supply vector and defined as $T_a^b\partial_b r+W\partial_a r$ with $T_{ab}$ representing the four-dimensional energy-momentum tensor within a two-dimensional sphere, and $V=\frac{4}{3}\pi r^3$ is it's volume. This equation is often referred to as the “unified first law” \cite{Hayward:1997jp,Mukohyama:1999sp,Hayward:1998ee}, which links the Misner-Sharp energy to the dynamics of black holes and cosmological horizons.

In the integration method, similar to Einstein's gravity, we can rewrite the field equations as follows
\begin{eqnarray} \label{eq003}
    dE_{eff} &=& A \Psi_a dx^a + W dV\nonumber\\
    &=&A(u,v) du + B(u,v) dv,
\end{eqnarray}
where the coefficients $A(u,v)$ and $B(u,v)$ can be calculated in terms of energy-momentum tensor components as 
\begin{widetext}
 \begin{eqnarray}\label{eq006}
A(u, v) &=& 4r^2 e^\varphi (r_u T_{uv} - r_v T_{uu})\nonumber \\
&=&\frac{1}{4 G} \Bigg[ 
f r^2 r_{,u} 
+ 2 e^{\varphi} \Big\{ r^2 [
r_{,u} \left( f_{R,uv} + f_R \varphi_{,uv} \right) 
+ r_{,v} \left( f_{R,u} \varphi_{,u} + f_{R,uu} \right) 
]\nonumber\\
&&
+ 2 r \Big[
r_{,u} \left( f_{R,v} r_{,u}- f_R r_{,uv} 
+ r_{,v} \left( f_{R,u} + f_R \varphi_{,u} \right) 
 \right) 
+ f_R r_{,v} r_{,uu} 
\Big]
\nonumber\\
&&+ 4 \Big[ 
2 e^{\varphi} r_{,v}^2 \left( 
r_{,u} f_{{\mathcal{G}},uu} + f_{{\mathcal{G}},u} \left( 
2 r_{,u} \varphi_{,u} + r_{,uu} 
\right) \right) 
+ r_{,v} \{ 
f_{{\mathcal{G}},u} \varphi_{,u} + f_{{\mathcal{G}},uu} \nonumber\\
&&
+ 2 e^{\varphi} r_{,u} [
f_{{\mathcal{G}},u} r_{,uv} + r_{,u} \left( 
\left( f_{{\mathcal{G}},v} + f_{\mathcal{G}} \varphi_{,v} \right) \varphi_{,u} 
+ f_{{\mathcal{G}},uv} + f_{\mathcal{G}} \varphi_{,uv} 
\right) + \left( f_{{\mathcal{G}},v} + f_{\mathcal{G}} \varphi_{,v} \right) r_{,uu} 
] \} 
\nonumber\\
&&+ r_{,u} \{ 
f_{{\mathcal{G}},uv} + f_{\mathcal{G}} \varphi_{,uv} 
+ 2 e^{\varphi} [ 
f_{{\mathcal{G}},v}  r_{,u} r_{,uv} 
+ f_{\mathcal{G}} \left( -r_{,uv}^2 + r_{,vv} \left( 
r_{,u} \varphi_{,u} + r_{,uu} 
\right) \right) 
] \} 
\Big] \Big\}\Bigg],
\notag\\
B(u, v) &=& 4r^2 e^\varphi (r_v T_{uv} - r_u T_{vv})\nonumber \\
&=&\frac{1}{4 G} \Bigg[ 
f r^2 r_{,v} 
+ 2 e^{\varphi} \Big\{
2 r
\Big[
r_{,v}^2 f_{R,u} 
+ f_R  r_{,vv} r_{,u} 
+ r_{,v} \left( \left( f_{R,v} + f_R\varphi_{,v} \right) r_{,u} 
- f_R r_{,uv} \right) 
\Big] 
\nonumber\\
&&+ 4 \Big[
\left( f_{{\mathcal{G}},v} \varphi_{,v} + f_{{\mathcal{G}},vv} \right) r_{,u} 
+ r_{,v} \left( f_{{\mathcal{G}},uv} + f_{\mathcal{G}} \varphi_{,uv} \right) 
\Big] 
+ r^2 \Big[
\left( f_{R,v}\varphi_{,v} + f_{R,vv} \right) r_{,u} 
+ r_{,v} \left( f_{R,uv} + f_R \varphi_{,uv} \right) 
\Big]\nonumber\\
&&
+ 16 e^{2 \varphi} \Big[
f_{{\mathcal{G}},v} r_{,vv} r_{,u}^2 
+ r_{,v}^2 \{ 
f_{{\mathcal{G}},u}r_{,uv} 
+ r_{,u} \left( f_{{\mathcal{G}},uv} + f_{\mathcal{G}}\varphi_{,uv} \right) 
+ \varphi_{,v} \left( 
r_{,u} \left( f_{{\mathcal{G}},u} + f_{\mathcal{G}}  \varphi_{,u} \right) 
+ f_{\mathcal{G}} r_{,uu} 
\right) \}\nonumber\\
&&
+ r_{,v} \{ 
\left( 2 f_{{\mathcal{G}},v}  \varphi_{,v} + f_{{\mathcal{G}},vv} \right) r_{,u}^2 
+ f_{{\mathcal{G}},v}  r_{,u} r_{,uv} 
- f_{\mathcal{G}}  r_{,uv}^2 
+ r_{,vv} \left( 
r_{,u} \left( f_{{\mathcal{G}},u} + f_{\mathcal{G}}\varphi_{,u} \right) 
+ f_{\mathcal{G}} r_{,uu} 
\right) 
\} 
\Big] 
\Big\} 
\Bigg],
\end{eqnarray}
In order to derive the generalized Misner-Sharp energy, one needs integrating Eq. (\ref{eq003}). Is integrability is provided by the condition 
\begin{eqnarray}
    \frac{\partial A(u,v)}{\partial v} = \frac{\partial B(u,v)}{\partial u}.
\end{eqnarray}
This Eq is not satisfied in general, unlike the pure Gauss-Bonnet gravity (see the explanation after Eq. (3.7) in \cite{Cai:2009qf}). Assuming that the integrability condition is satisfied, one obtains the Misner-Sharp energy  
\begin{eqnarray}\label{eq004}
E_{\text{eff}} &=& \int A(u,v) du + \int\Big[ B(u,v) -\frac{\partial}{\partial v} \int A(u,v) du \Big] dv\nonumber \\
&=&\frac{r}{2G}\Big[\Big(\left(1+2e^\varphi r_{,u}r_{,v}\right)f_{R}+\frac{1}{6}r^2\left(f-f_R R\right)+re^\varphi \left(f_{R,u}r_{,v}+f_{R,v}r_{,u}\right)\Big)+2r_{,v}e^{2\varphi}\Big(\left(e^{-\varphi} +2r_{,v}r_{,u}\right)f_{{\mathcal{G}},u}+2r^2_{,u} f_{{\mathcal{G}},v}\Big)\Big]\nonumber\\
&&-\frac{1}{2G}\int \Bigg[ f_{R,u}e^\varphi (r^2 r_{,v})_{,u}+f_{R,v}r^2 (r_{,u}e^{2 \varphi})_{,u}+f_{R,u}\Big(r-\frac{1}{6}r^3 R\Big)\nonumber\\
&& \;\;\;\;\;\;\;\;\;\;\;\;\;\;\;\;\;\;+4e^\varphi r^2_{,u} (\frac{f_{{\mathcal{G}},u}}{r_{,u}})_{,v}+4f_{{\mathcal{G}},v}r_{,v}( r^2_{,u} e^{2\varphi})_{,u}+8e^{2\varphi} r_{,v} r_{u,} r_{,uv} f_{{\mathcal{G}},u}\Bigg]du,
\end{eqnarray}
where we used
\begin{eqnarray} 
f_{,u}&=&f_R R_{,u} +f_{\mathcal{G}} \mathcal{G}_{,u} ,\nonumber \\
    R&=&2[\frac{1}{r^2}+e^\varphi(2\frac{r_{,v}r_{,u}}{r^2} -\varphi_{,uv}+4\frac{r_{,uv}}{r})]\nonumber,\\
    \mathcal{G} &=& -\frac{8e^\varphi}{r^2}[-2e^\varphi r^2_{,uv}+\varphi_{,uv}+2e^\varphi r_{,vv}(r_{,u}\varphi_{,u}+r_{,uu})+2e^\varphi r_{,v}\left(r_{,u}\varphi_{,uv}+\varphi_{,v}\left(r_{,u}\varphi_{,u}+r_{,uu}\right)\right)].
\end{eqnarray}
We can see that equation (\ref{eq004}) reduces to the Misner-Sharp energy in Einstein gravity if $f_R=1 $ and $f_{\mathcal{G}}=0$ and it reduces to Eq. (3.8) in \cite{Cai:2009qf} for $f_{G} = 0$. If\textit{ A(u,v)} and \textit{B(u,v)} do not  satisfy the integrability condition, one cannot obtain the Misner-Sharp energy as above.
\end{widetext}
\subsection{Conserved charge method}
Here, we  proceed to drive the Misner-Sharp energy using the conserved charge method.\\
In spherically symmetric space-times, the concept of a Kodama vector plays a pivotal role in defining conserved quantities \cite{Abreu:2010ru,Dorau:2024zyi}. Unlike the Killing vector, the Kodama vector remains well defined in dynamic spacetimes and facilitates the construction of a conserved current. In Einstein’s General Relativity, the combination of the energy-momentum tensor with the Kodama vector generates a conserved current, leading to the definition of the Misner-Sharp energy, which is a quasi-local measure of energy within a given region. The Misner-Sharp energy has proven to be a valuable tool in understanding energy content, including matter, radiation, and gravitational contributions.
In extended theories of gravity, such as Gauss-Bonnet gravity, Maeda and Nozawa \cite{Maeda:2007uu} utilized the Kodama vector and conserved current methods to derive a generalized form of the Misner-Sharp energy. This generalized energy not only retains many properties of the original Misner-Sharp energy, such as monotonicity and positivity, but also reflects the contributions from the higher-order curvature terms inherent in the theory. Inspired by this approach, our aim is to extend the derivation of the Misner-Sharp energy to the $f(R,\mathcal{G})$ gravity framework. By constructing a conserved current using a generalized Kodama vector, we demonstrate that the quasi-local mass defined in this framework can be naturally interpreted as the Misner-Sharp energy counterpart in $f(R,\mathcal{G})$ gravity. By adopting this method, we provide a consistent and physically meaningful extension of the Misner-Sharp energy to $f(R,\mathcal{G})$ gravity, offering insight into the energy distribution and thermodynamics of this modified theory of gravity. This analysis not only broadens the applicability of the Misner-Sharp energy but also deepens our understanding of quasi-local energy in the context of higher-order gravitational theories.\\
The Kodama vector is defined as \cite{Kodama:1979vn,Minamitsuji:2003at}
\begin{eqnarray}
    K^\mu = - \epsilon^{\mu \nu }\nabla_\nu r,
\end{eqnarray}
Here,  $\epsilon^{\mu\nu} = \epsilon^{ab}(dx^a)_\mu(dx^b)_\nu$,  and $\epsilon^{ab}$ represents the volume element of the squared mass parameter ($M^2$), which is used as a measure of mass in these theories. This quantity is also related to the two-dimensional metric tensor ($h^{ab}$) in the subspace perpendicular to symmetric spheres, typically defined in double-null coordinates. By using the verification of the identity and the Bianchi relations below, it can be easily checked that the conservation of the energy-momentum tensor $T_{\mu\nu}$ holds in (\ref{eq002}) 
\begin{eqnarray}
  && \left(\Box\nabla_\nu - \nabla_\nu \Box\right)A = R_{\mu\nu}\nabla^\mu A,\nonumber\\
  && \nabla^\mu R_{\mu\nu} = \frac{1}{2} \nabla_\nu R,
   \end{eqnarray}
where $A$ is an arbitrary scalar function. The definition of the energy current is given by
\begin{eqnarray}\label{eq002.1}
    J^\mu = -T^\mu_\nu K^\nu.
\end{eqnarray}
The energy follow (\ref{eq002.1}) in $f(R,\mathcal{G})$ gravity becomes divergent free, i.e. $\nabla_{\mu}J^{\mu}=0$, under the following condition
   \begin{eqnarray}
   &&\Big(\nabla_\mu \nabla_\nu f_R  +2 R \nabla_\mu \nabla_\nu f_{\mathcal{G}} - 4 R_\mu^\lambda \nabla_\lambda \nabla_\nu f_{\mathcal{G}}\nonumber\\
   &&- 4 R_\nu^\lambda \nabla_\lambda \nabla_\mu f_{\mathcal{G}} +R_{\mu\alpha\beta\nu} \nabla^\alpha
\nabla^\beta f_{\mathcal{G}}  \Big)\nabla^\mu K^\nu = 0.
   \end{eqnarray}\\
    Under this condition, the corresponding conserved charge will be 
\begin{eqnarray}
    Q_J = \int_\Sigma J^\mu d\Sigma_\mu,
\end{eqnarray}
where $\Sigma$ represents a specific hypersurface, while $d\Sigma_\mu$ is defined as the directed surface line element on $\Sigma$, expressed through the formula $ d\Sigma_\mu = \sqrt{-g} dx^{\nu}dx^{\lambda} dx^{\rho} \delta_{\mu \nu \lambda \rho}$, where $g$ denotes the metric determinant, and $\delta_{\mu \nu \lambda \rho}$ is the antisymmetric tensor associated with the coordinates $x^{\nu}$, $x^{\lambda}$, and $x^{\rho}$.

\begin{widetext}
Using the metric (\ref{eq001}) and equations in (\ref{eq008}), we obtain the conserved charge $Q_J$ as
    \begin{eqnarray}\label{eq0004}
        \hspace{-4cm}Q_J &=&\int_\Sigma J^\mu d\Sigma_\mu\nonumber \\
        &=&\frac{r}{2G}\Big[\Big(\left(1+2e^\varphi r_{,u}r_{,v}\right)f_{R}+\frac{1}{6}r^2\left(f-f_R R\right)+re^\varphi \left(f_{R,u}r_{,v}+f_{R,v}r_{,u}\right)\Big)+2r_{,v}e^{2\varphi}\Big(\left(e^{-\varphi} +2r_{,v}r_{,u}\right)f_{{\mathcal{G}},u}+2r^2_{,u} f_{{\mathcal{G}},v}\Big)\Big]\nonumber\\
&&-\frac{1}{2G}\int \Bigg[ f_{R,u}e^\varphi (r^2 r_{,v})_{,u}+f_{R,v}r^2 (r_{,u}e^{2 \varphi})_{,u}+f_{R,u}\Big(r-\frac{1}{6}r^3 R\Big)\nonumber+4e^\varphi r^2_{,u} (\frac{f_{{\mathcal{G}},u}}{r_{,u}})_{,v}+4f_{{\mathcal{G}},v}r_{,v}( r^2_{,u} e^{2\varphi})_{,u} \nonumber \\
&&+8e^{2\varphi} r_{,v} r_{u,} r_{,uv} f_{{\mathcal{G}},u}\Bigg]du. 
    \end{eqnarray}
    By comparing equations (\ref{eq004}) and (\ref{eq0004}), we observe that both methods yield identical results.
\section{GENERALIZED MISNER-SHARP ENERGY IN $f(R,\mathcal{G})$ THEORY: SPECIAL CASES}
In this section, we derive the generalized Misner-Sharp energy in $f(R, \mathcal{G})$ gravity for two distinct spacetimes: Static spherically symmetric and dynamic FLRW spacetime \cite{Seifert:2007fr}. Using the integration method, we obtain explicit expressions for the energy in both cases, highlighting the contributions of curvature corrections from the Ricci scalar and the Gauss-Bonnet term. For the static case, we focus on simplifying the field equations to isolate the energy distribution, enabling comparisons with standard results in Einstein's general relativity, such as the Schwarzschild solution \cite{Virbhadra:1999nm}. In the FLRW case, we explore how the energy relates to the total matter content within a specified radius.
 
\subsection{Generalized Misner-sharp energy in static spherically symmetric case}
Here, we consider the static spherically symmetric spacetime with the line element 

\begin{eqnarray}
    ds^2=-\lambda(r)dt^2+g(r)dr^2+r^2d\Omega^2_2,
\end{eqnarray}
where $\lambda(r)$ and $g(r)$ are arbitrary functions.
In order to use the equation (\ref{eq003}), we first obtain $A(r)$ and $B(r)$ using energy-momentum tensor components as
\begin{eqnarray}
A(r)&=&\frac{4\pi r^2}{g}T_{tr}=0,\nonumber\\
B(r)&=&\frac{4\pi r^2}{\lambda}T_{tt}\nonumber\\
  &=&\frac{1}{8  G  g^3 \lambda^2} \Big[ 
2 f r^2 g^3 \lambda^2 
- g \{ 4 f_{\mathcal{G}} (g-1 ) + r^2 gf_R\} \lambda'^2 
+ \lambda \{ 
[4 r f_R  g^2 - r g'-4 f_{\mathcal{G}} (g-3) g'  ]\lambda'\nonumber\\
&&
+ 2 g [ 4 f_{\mathcal{G}} (g-1) + r^2 f_R  g ] \lambda'' \}+ 2 \lambda^2 \{
 g [( 4 f_{\mathcal{G}}' + r^2 f_R' ) g' + 8 f_{\mathcal{G}}'' - 2 g \left( 2 r f_R' + 4 f_{\mathcal{G}}'' + r^2 f_R'' \right) ] \nonumber\\
 &&-12g'f_{\mathcal{G}}'\}\Big],
\end{eqnarray}
Here, a prime indicates differentiation with respect to 
 coordinate $r$. With the integrability condition satisfied, the Misner-Sharp energy is obtained as
\begin{eqnarray}\label{eq007}
E_{eff}=\int B(r)dr&=&\frac{r}{2G}\Big[(1-h^{ab}\partial_ar \partial_br)f_R+\frac{r^2}{6}(f-Rf_R-\mathcal{G}f_{\mathcal{G}})-rh^{ab}\partial_af_R \partial_br-4\frac{f'_{\mathcal{G}}}{rg}(1-\frac{1}{g})\Big]\nonumber\\
        &&-\frac{1}{2G}\int\Big[(r^2\frac{g'}{2g^2}+r-\frac{r}{g}-\frac{1}{6}r^3R)f_{R,r}+(2\frac{g'}{g^2}-2\frac{g'}{g^3}-\frac{1}{6}r^3\mathcal{G})f_{\mathcal{G},r}\Big]dr,
    \end{eqnarray}
\end{widetext}
where $\partial f_{R}/\partial r = f_{R,r}$ and $\partial f_{\mathcal{G}}/\partial r = f_{\mathcal{G},r}$. One observers that under conditions ($i$) $f_{R,r} = 0$ or $\frac{r^2 g'}{2g^2}+r-\frac{r}{g}-\frac{r^3}{6}R = 0$, and ($ii$) $f_{\mathcal{G},r}=0$ or $\frac{2g'}{g^2}-\frac{2g'}{g^3}-\frac{1}{6}\mathcal{G} = 0 $, the integral term in equation (\ref{eq007}) will be eliminated. In a particular, case where $f(R, \mathcal{G})=R$, our result corresponds to \cite{Cai:2009qf}.

\subsection{Generalized Misner-sharp energy in FLRW spacetime}
Here, we obtain the Misner-Sharp energy for the metric with the line element 
\begin{eqnarray}
ds^2 &=& -dt^2 +e^{2\psi(t,\rho)}d\rho^2\nonumber\\
&&+r^2(t,\rho)\left(d\theta^2 +\sin^2 \theta d\phi^2 
\right).
\end{eqnarray}
Considering $ r( t,\rho) = a(t)\rho$ and $e^{\psi(t,\rho)}=\frac{a(t)}{\sqrt{1-k\rho^2}}$, it reduces to FLRW line element. To be continued, we rewrite the equation (\ref{eq003}) in the new variables
\begin{eqnarray}
dE_{eff}=A(r,\rho)dt+B(r,\rho)d\rho.
\end{eqnarray}
where $A(t,\rho)$ and $B(t,\rho)$ are obtained as
\begin{eqnarray}
        A(r,\rho) &=& 4\pi r^2e^{-2\psi}(T_{t\rho}r_{,\rho}-T_{\rho\rho}r_{,t}),\nonumber \\
        B(r,\rho) &=&4 \pi r^2(T_{tt}r_{,\rho}-T_{t\rho}r_{,t}).
    \end{eqnarray}
Here, assuming the satisfaction of the integrability  condition 
$\partial A(t,\rho)/\partial\rho = \partial B(t,\rho)/\partial t$, one finds the Misner-Sharp energy as follows
\begin{widetext}
\begin{eqnarray}\label{eq0005}
E_{eff} &=& \int B(t,\rho)d\rho + \int \Big[A(t,\rho)-\frac{\partial }{\partial t} \int B(t,\rho)d\rho\Big] d\rho \nonumber \\
&=& \frac{1}{2G}\Big[r(1- h^{ab}\partial_a r\partial_b r)f_R+\frac{r^3}{6}(f-f_RR-f_{\mathcal{G}}\mathcal{G})-r^2h^{ab}\partial_a f_R\partial_b r\nonumber \\
&&+4f_{\mathcal{G},t}r_{,t}(r^2_{,t}-e^{-2\psi}r^2_{,\rho})+4e^{-2\psi} r_{,\rho}f_{\mathcal{G},\rho}(e^{-2\psi}r^2_{,\rho}-r^2_{,t}-1)\Big]\nonumber\\
&&\frac{1}{2G}\int\Big\{f_{R,\rho}\Big[(-e^{-2\psi}r^2r_{,\rho}\psi_{,\rho}+e^{-2\psi}r^2r_{,\rho\rho}-r^2r_{,t}\psi_{,t})-r(1+r^2_{,t}-e^{-2\psi}r^2_{,\rho})+\frac{1}{6}r^3R\Big]+r^2f_{R,t}(\psi_{,t}r_{,\rho}-r_{,t\rho})\nonumber\\
&&+f_{\mathcal{G},\rho}e^{-2\psi}\Big[r^2_{,\rho}\left(e^{-2\psi}(r_{,\rho}\psi_{,\rho}-r_{,\rho\rho})+r_{,t}\psi_{,t}\right)+(r^2_{,t}+1)\left(r_{,\rho\rho}-r_{,\rho}\psi_{,\rho}-e^{2\psi}r_{,t}\psi_{,t}\right)\Big]\nonumber \\
&&+f_{\mathcal{G},t}e^{-2\psi}\Big[r^2_{,\rho}r_{,t\rho}+e^{2\psi}r_{,t}(1-r_{,t}r_{,t\rho})+r_{,\rho}\psi_{,t}\left(e^{2\psi}(r^2_{,t}-1) -r^2_{,\rho}\right)\Big]\Big\}d\rho.
\end{eqnarray}
One observe that the expression in (\ref{eq0005}) reduces to Misner-Sharp energy in GR, with $f(R,\mathcal{G}) = R$ as all the terms expect the first term vanish.\\

\end{widetext}

\section{ APPARENT HORIZON THERMODYNAMICS IN FLRW SPACETIME}
At this point, we briefly discuss the thermodynamics of the FLRW spacetime for a class of $f(R,\mathcal{G})$. Consider the line element for a four-dimensional homogeneous and isotropic universe as
\begin{eqnarray}\label{eq401}
    ds^2=-dt^2+a^2(t)\left[\frac{dr^2}{1-kr^2}+r^2d\Omega^2_{2}\right],
\end{eqnarray}
where, $k$ represents the spatial curvature. For simplicity, we shall denote $a(t)\equiv a$. The dynamical apparent horizon for FLRW metric is given by \cite{Cai:2005ra}
\begin{eqnarray}\label{eq402}
    R_A = \frac{1}{\sqrt{H^2+k/a^2}},
\end{eqnarray}
where $H\equiv \dot{a}/a$ is the Hubble parameter. The surface gravity at the apparent horizon is \cite{Cai:2005ra,Abdusattar:2024sgk}
\begin{eqnarray}
    \kappa = -\frac{1}{R_A}\left( 1-\frac{\dot{R_A}}{2 H R_A}\right),
\end{eqnarray}
and consequently,  the Hawking temperature, defined in terms of surface gravity, reads as \cite{Abdusattar:2024sgk}
\begin{eqnarray}\label{eq405}
   T=\frac{\lvert\kappa\rvert}{2\pi}.
\end{eqnarray}
The thermodynamic pressure is defined by the fluid's work density \cite{Hayward:1993wb,Hayward:1994bu,Akbar:2006kj}  
\begin{eqnarray}\label{eq400}
    P\equiv W = -\frac{1}{2} h_{ij}T^{ij}.
\end{eqnarray}
In the following, we consider a particular case of $f(R,\mathcal{G})$ theory as \cite{Nojiri:2005jg,Odintsov:2018nch,Abbas:2014vka} 
\begin{eqnarray}\label{eqf(R,G)}
    f(R,\mathcal{G})=R+f_0 \mathcal{G}^n,
\end{eqnarray}
where $n$ is an even positive number and $f_0$ is an arbitrary constant. \\
\begin{widetext}
Hence, the Misner-sharp energy within the apparent horizon can be obtained as 
\begin{eqnarray}\label{eq432}
    E_{eff}&=&\frac{R^3_A}{12G}\Bigg[\frac{6}{R_A^2}+f_0(n-1)\mathcal{G}^{n-2}\Big( -\mathcal{G}^2+ \frac{12n\dot{R_A}\dot{\mathcal{G}}}{R_A^3(1-2 \pi R_A T)}
 \Big)\Bigg],
\end{eqnarray}
where the Gauss-Bonnet scalar reads
\begin{eqnarray}
    \mathcal{G} = \frac{24}{R_A^4}(4\pi R_A T-1).
\end{eqnarray}
From Eqs. (\ref{eq000}), (\ref{eq401}),  (\ref{eq402}) and (\ref{eq405}), one can express density $\rho$ and pressure $p$ in terms of $R_A$ and its derivatives as
\begin{eqnarray}\label{eq403}
    \rho&=&\frac{1}{16\pi G R_A^3 \mathcal{G}^2}
    \Big( f_0(n-1)[R_A^3 \mathcal{G}^2+\frac{12 n \dot{R_A}\dot{\mathcal{G}}}{2\pi R_A T -1}]\mathcal{G}^n-6 R_A \mathcal{G}^2 \Big),\nonumber \\
p&=&\frac{1}{16\pi G R_A^3 \mathcal{G}^3}
    \Big( 2 (1-8\pi R_A T) R_A\mathcal{G}^3+f_0(n-1)R_A^3 \mathcal{G}^{n+3}-8f_0 (n-2)(n-1)nR_A \mathcal{G}^n \dot{\mathcal{G}^2} \nonumber\\
    &&- 8 f_0 (n-1)n\mathcal{G}^{n+1}\Big[\frac{\dot{R_A}(1-4\pi R_A T)\dot{\mathcal{G}}}{2\pi R_A T -1} + R_A \ddot{\mathcal{G}}\Big] \Big).
\end{eqnarray}
Assuming that the variation of $R_A$ with respect to time is small, the surface gravity at the apparent horizon of the FLRW universe becomes negative $(\kappa<0)$ \cite{Wu:2020fij,Hayward:1993wb,Dolan:2013ft,Abdusattar:2022bpg}. This imposes a constraint on the perfect fluid within the framework of $f(R,\mathcal{G})$ gravity, which is expressed as follows
\begin{eqnarray}
    \frac{\dot{R_A}}{2HR_A}<1,
\end{eqnarray}
or equivalently
\begin{eqnarray}
    TR_A>0.
\end{eqnarray}
Using the thermodynamic pressure at the apparent horizon of FLRW universe (\ref{eq400}), we find 

\begin{eqnarray}\label{eq407}
    P &=& \frac{1}{4\pi GR_A^3 \mathcal{G}^3} \Big(2R_A (2\pi R_A T-1)\mathcal{G}^3 + 2f_0(n-2)(n-1)nR_A \mathcal{G}^n \dot{\mathcal{G}}^2 \nonumber\\
    &&\;\;\;\;\;\;\;\;\;\;\;\;\;\;\;\;\;\;+ \frac{f_0 n(n-1)\mathcal{G}^{n+1}}{2\pi R_A T - 1}[(5-8 \pi R_A T) \dot{R_A} \dot{\mathcal{G}} +2R_A (2\pi R_A T - 1 )\ddot{\mathcal{G}}]\Big).
\end{eqnarray}
By using Eq. (\ref{eq407}), we determine the critical points by applying the conditions
\begin{eqnarray}
    \left(\frac{\partial P}{\partial V}\right)_T = \left(\frac{\partial^2 P}{\partial V^2}\right)_T = 0,
\end{eqnarray}
or equivalently
\begin{eqnarray}\label{eq408}
    \left(\frac{\partial P}{\partial R_A}\right)_T = \left(\frac{\partial^2 P}{\partial R_A^2}\right)_T = 0.
\end{eqnarray}
These conditions, as explored in \cite{Bena:2012hf,Hu:2018qsy,Arcadi:2021mag,Abdusattar:2023hlj}, allow us to identify the critical points of the thermodynamic system.\\
To further analyze the thermodynamic properties of the universe, we compare $dE = -TdS + WdV$ with the standard form of first law of thermodynamic as
\begin{eqnarray}
    U := -E.
\end{eqnarray}
Consequently, the enthalpy $\mathcal{H}$ is obtained as
\begin{eqnarray}
    \mathcal{H} = -E_{eff} + PV,
\end{eqnarray}
where using $V = \frac{4\pi R_A^3}{3}$ and equations (\ref{eq407}) and (\ref{eq432}), it reads as
\begin{eqnarray}
    \mathcal{H}&=&\frac{1}{12G}\Big( 2R_A(8\pi R_A T-7)+f_0(n-1)R_A^3\mathcal{G}^n + 8f_0 (n-2)(n-1)nR_A \mathcal{G}^{n-3}\dot{\mathcal{G}}^2 \nonumber\\
    &&+ \frac{4f_0 (n-1)n\mathcal{G}^{n-2}}{2 \pi R_A T -1}\Big[ (\dot{R_A}(5-8 \pi R_A T )+12R_A^2(1-2\pi R_A T)^2)\dot{\mathcal{G}} + 2R_A (2\pi R_A T -1)\ddot{\mathcal{G}} \Big]  \Big).
\end{eqnarray}
The specific heat at constant pressure $C_P$ serves as a key indicator of the thermodynamic stability of the universe, showing how the system’s energy changes with temperature at a constant pressure. A positive $C_P$ indicates thermodynamic stability, while a negative $C_P$ suggests instability. $C_P$ can be obtained as
\begin{eqnarray}
    &&C_P =\left(\frac{\partial\mathcal{H}}{\partial T}\right)_P =-\frac{\pi R_A(2R_A \alpha^2 \mathcal{G}^2 - f_0 (n-1)n\dot{R_A}\mathcal{G}^n\dot{\mathcal{G}})\beta}{6G \alpha^2 \mathcal{G}^2\gamma},
\end{eqnarray}
where
\begin{eqnarray}
\alpha&\equiv&1-2\pi R_A T,\nonumber\\
\beta&\equiv&2R_A(1-8\alpha)\alpha^2\mathcal{G}^3 + 3f_0 (n-1)R_A^3\alpha^2 \mathcal{G}^{n+3} + 8f_0 (n-2) (n-1)nR_A \alpha^2 \mathcal{G}^n\dot{\mathcal{G}}^2\nonumber\\
&&+ 4f_0 (n-1)n\mathcal{G}^{n+1}[(\dot{R_A}(\alpha-1) + 12R_A^2(1-3\alpha)\alpha^2)\dot{\mathcal{G}}+2R_A \alpha^2 \ddot{\mathcal{G}}],\nonumber\\
\gamma&\equiv&2R_A\alpha^2(1+\alpha)\mathcal{G}^3 - 4f_0n(2-3n-n^2)R_A \alpha^2 \mathcal{G}^n\dot{\mathcal{G}}^2 + f_0 (n-1)n\mathcal{G}^{n+1}[\dot{R_A}(12\alpha^2+4\alpha-1)\dot{\mathcal{G}} - 4R_A \alpha^2 \ddot{\mathcal{G}}],
\end{eqnarray}
Varying $n$ allows for the analysis of different stability states of the universe.
The specific heat at a constant volume $C_V$ measures the change in the internal energy with the temperature at a constant volume, playing a crucial role in understanding the universe’s behavior in adiabatic processes. It can be evaluted as
\begin{eqnarray}
    C_V &=&  C_P-T\left(\frac{\partial P}{\partial T}\right)_V\left( \frac{\partial V}{\partial T} \right)_P\nonumber\\
    &=&-\frac{\pi R_A(2R_A \alpha^2 \mathcal{G}^2 - f_0 (n-1)n\dot{R_A}\mathcal{G}^n\dot{\mathcal{G}})}{6G \alpha^2 \mathcal{G}^2\gamma}\left( \beta +12(\alpha-1)\mathcal{G}[2R_A \alpha^2 \mathcal{G}^2 -f_0 \dot{R_A}\dot{\mathcal{G}}\mathcal{G}^nn(n-1)]\right).
\end{eqnarray}
The adiabatic index $q$, defined as the ratio of specific heats, describes the universe’s behavior in adiabatic processes
\begin{eqnarray}\label{eq q}
    q = -\frac{C_p}{C_v} = -\frac{\beta}{12\mathcal{G}(\alpha-1)[2R_A\mathcal{G}^2\alpha^2-f_0\dot{R_A}\mathcal{G}^n\dot{\mathcal{G}}n(n-1)]+\beta}.
\end{eqnarray}
Depending on $q$ values there are three possible cases: 

\begin{itemize}
    \item A static universe when \( q = 0 \),
    \item A decelerating universe when \( q > 0 \),
    \item An accelerating universe when \( q < 0 \).
\end{itemize}
To investigate the cosmological evolution within the framework of $f(R,\mathcal{G})$ theory in (31), we adopt a specific form for the scale factor
\begin{eqnarray}\label{a(t)}
    a(t)=N(t+hn)^n,
\end{eqnarray}
where $N$ is a normalization constant, $h$ is a temporal shift parameter, and $n$ is a positive number governing the behavior of cosmological evolution. This form, inspired by Ref. \cite{Odintsov:2018nch}, is well-suited for describing quasi-de Sitter inflationary evolution and is compatible with slow-roll conditions. Its power-law structure facilitates analytical solutions to the Friedmann equations, enabling the study of cosmological dynamics across various curvature regimes. For a flat universe ($k = 0$), the behavior of $q(t)$ for $n = 2, 3, 4$ is plotted in Fig. (\ref{fig1}) 
\begin{figure}[H]
    \centering
    \includegraphics[width=0.7\linewidth]{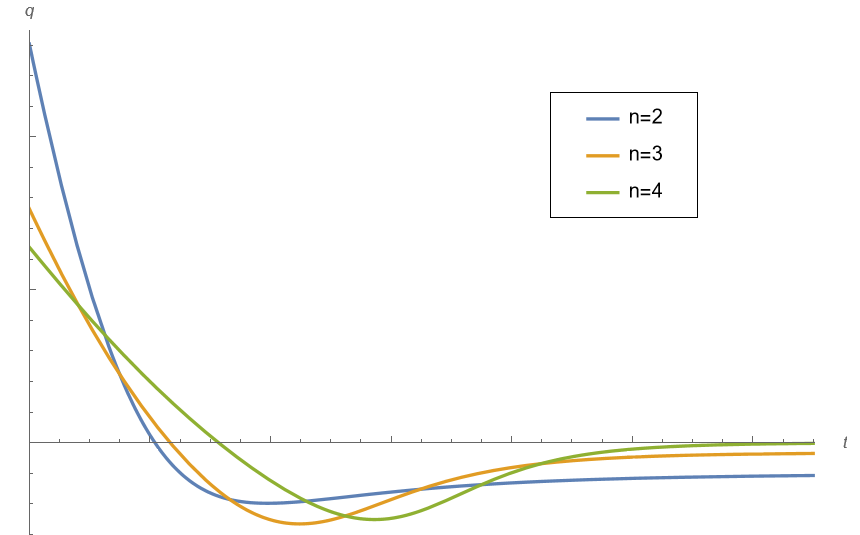}
    \caption{Adiabatic index $q(t)$ for $n = 2, 3, 4$ in a flat universe ($k = 0$) in the $f(R, \mathcal{G})$ theory, showing the transition from a decelerating to accelerating phase.}
    \label{fig1}
\end{figure}
The figure illustrates that $q$ initially exhibits positive values, corresponding to a decelerating phase of cosmic expansion.  Over time, $q$ under goes a phase transitions toward negative values, indicating an accelerating expansion phase. For $n = 2$ and $n = 3$, $q$ remains negative, signifying sustained accelerated expansion, with the transition occurring later for $n = 3$ compared to $n = 2$. For $n = 4$, $q$ approaches zero at late times, corresponding to a static universe. For $n > 4$, after the accelerating phase, $q$ returns to positive values, indicating a stable decelerating phase. These behaviors are driven by the higher-order curvature terms $\mathcal{G}^n$ in the $f(R, \mathcal{G})$ theory, which play a significant role in shaping cosmological evolution.
\end{widetext}
\section{CONCLUSION AND DISCUSSION}
In this study, we explore the Misner-Sharp energy, a key concept in gravitational physics, within the framework of $f(R, \mathcal{G})$ gravity, which extends Einstein's general relativity by incorporating the Ricci scalar $R$ and the Gauss-Bonnet scalar $\mathcal{G}$. Our aim is to generalize this quasilocal energy concept and examine its implications across various contexts, from static black holes to a dynamic universe, while connecting it to cosmic thermodynamics. We employ two distinct methods to achieve this. First, through an integration technique, we derive explicit expressions for the Misner-Sharp energy in three scenarios: a general spherically symmetric spacetime, a static spherically symmetric case, and the dynamic FLRW spacetime. In each case, the energy includes additional terms arising from the curvature corrections of $f(R, \mathcal{G})$ gravity. These results reduce to standard Einstein gravity when the modifications are set to zero ($f_R = 1, f_{\mathcal{G}} = 0$). In the FLRW case, the energy corresponds elegantly to the total matter content within a specified radius, linking geometric properties to cosmic matter distribution. Our second approach uses the conserved charge method, leveraging the Kodama vector, a powerful tool for dynamic spacetimes where conventional symmetries may fail. This method yields an alternative derivation of the Misner-Sharp energy, which aligns with the integration method results under conditions such as integrability and the absence of energy divergence. The agreement between both methods confirms the robustness of our findings, demonstrating that the energy in $f(R, \mathcal{G})$ gravity captures both matter and the complex interplay of spacetime curvature.

We also analyze the thermodynamics of this model, focusing on the apparent horizon of the FLRW universe. Using a specific class of $f(R, \mathcal{G})$ models, defined as $f(R, \mathcal{G}) = R + f_0 \mathcal{G}^n$, we calculate key quantities such as energy density, pressure, temperature, and Misner-Sharp energy within this horizon. The curvature terms introduce a non-equilibrium aspect to the thermodynamics, with negative surface gravity under certain conditions, suggesting that $f(R, \mathcal{G})$ gravity can mimic effects typically attributed to exotic fluids without invoking dark energy. Notably, the adiabatic index $q(t)$, as shown in Fig. (\ref{fig1}), reveals a phase transition from a decelerating to an accelerating universe for $n=2$ and $n=3$, with sustained acceleration, while $n=4$ approaches a static universe, and $n>4$ indicates a return to a stable decelerating phase.

\nocite{*}

\bibliography{apssamp}% Produces the bibliography via BibTeX.
%146 citations counted in INSPIRE as of 03 Dec 2024

\end{document}